\documentclass[12pt,oneside, a4paper]{article}

\pdfoutput=1  
 
\ifx\pdfoutput\undefined
\usepackage[dvips,bookmarks=false]{hyperref}	
\else
\usepackage{hyperref}	
\fi
\hypersetup{colorlinks,bookmarksopen,bookmarksnumbered,citecolor=blue,linkcolor=blue,pdfstartview=FitH,urlcolor=blue}

\usepackage{graphicx}
\usepackage{subfigure}
\usepackage{amssymb}
\usepackage{cite}
\usepackage{bm}
\usepackage{amsmath,amsthm}
\usepackage{cleveref}
\usepackage{siunitx}
\usepackage{slashed}
\usepackage{multirow}
\usepackage{multicol}
\numberwithin{equation}{section}

\usepackage[normalem]{ulem}

\newcommand{\capdef}{}
\newcommand{\mycaption}[2][\capdef]{\renewcommand{\capdef}{#2}%
       \caption[#1]{{\footnotesize #2}}}

\begin{document}

\begin{titlepage}

\begin{center}

\vspace*{2cm}
        {\Large\bf Model-independent search for T violation\\[2mm]
        with T2HK and DUNE}
\vspace{1cm}

\renewcommand{\thefootnote}{\fnsymbol{footnote}}
{\bf Sabya Sachi Chatterjee}$^a$\footnote[1]{sabya.chatterjee@kit.edu},
{\bf Sudhanwa Patra}$^b$\footnote[2]{sudhanwa@iitbhilai.ac.in}
{\bf Thomas Schwetz}$^a$\footnote[3]{schwetz@kit.edu},\\
{\bf Kiran Sharma}$^{a,b}$\footnote[4]{kirans@iitbhilai.ac.in}
\vspace{5mm}

$^a$ {\it {Institut f\"ur Astroteilchenphysik, Karlsruher Institut f\"ur Technologie (KIT),\\ 76021 Karlsruhe, Germany}}\\
$^b$ {\it {Department of Physics, Indian Institute of Technology Bhilai, Durg-491002, India}}

\vspace{8mm} 

\today
  
\vspace{8mm} 

\abstract{We consider the time reversal (T) transformation in neutrino oscillations in a model-independent way by comparing the observed transition probabilities at two different baselines at the same neutrino energy. 
We show that, under modest model assumptions, if the transition probability $P_{\nu_\mu\to\nu_e}$ around $E_\nu \simeq 0.86$~GeV measured at DUNE is smaller than the one at T2HK the T symmetry has to be violated. Experimental requirements needed to achieve good sensitivity to this test for T violation are  to obtain enough statistics at DUNE for $E_\nu \lesssim 1$~GeV (around the 2nd oscillation maximum), good energy resolution (better than 10\%), and near-detector measurements with a precision of order 1\% or better. 
}

\end{center}

\end{titlepage}

\renewcommand{\thefootnote}{\arabic{footnote}}
\setcounter{footnote}{0}

\setcounter{page}{2}
\tableofcontents

\section{Introduction}

One of the top priorities in current neutrino oscillation research is
the determination of leptonic CP violation, which thanks to the CPT
theorem is directly related to T violation. The presence of this
phenomenon in neutrino oscillations has been pointed out long
ago~\cite{Cabibbo:1977nk,Bilenky:1980cx,Barger:1980jm}. From the
operational point of view the usual search method is rather
model-dependent: one assumes a particular model for neutrino mass,
mixing and interactions and then performs a model-dependent fit to the
observed event spectra, and checks if the fit shows preference for CP
violating complex phases in the model. Indeed, in the standard
three-flavour model, this amounts to determining a single complex
phase in the lepton mixing
matrix~\cite{Pontecorvo:1967fh,Gribov:1968kq,Maki:1962mu}, the
so-called Dirac phase $\delta_{\rm CP}$ \cite{Kobayashi:1973fv}.

This method is intrinsically model-dependent, as it does not allow to construct model-independent measures of (intrinsic) CP violation. One challenge in this respect is the presence of matter effects~\cite{Wolfenstein:1977ue} which induces enviromental CP violation and obscures fundamental (or intrinsic) CP violation of the theory~\cite{Bernabeu:2018use}. In contrast, T violation provides in principle a cleaner signature, as the matter effect does not introduce environmental T violation if the
fundamental theory is T invariant, as long as the matter density is symmetric between the source and the detector~\cite{Krastev:1988yu,Akhmedov:2001kd}. An incomplete list of papers on T violation in
neutrino oscillations is~\cite{Cabibbo:1977nk,Kuo:1987km,Krastev:1988yu,Toshev:1989vz,Toshev:1991ku,Arafune:1996bt,Parke:2000hu,Akhmedov:2001kd,Schwetz:2007py,Xing:2013uxa,Petcov:2018zka,Bernabeu:2019npc}. 

Applying the time reversal transformation to the oscillation probability, one finds
\begin{equation}\label{eq:T}
\mathcal{T}[P_{\alpha\to\beta}(L)] = P_{\beta\to\alpha}(L) \,.  
\end{equation}
Constructing T asymmetric observables based on this property is challenging, 
as it amounts to an interchange of neutrino flavours of source and detector.
Modern long-baseline experiments have good sensitivity to the 
$P_{\nu_\mu\to\nu_e}$ and $P_{\bar\nu_\mu\to\bar\nu_e}$ appearance channels, but it is much more difficult to search for the T reversed transitions, due to experimental obstacles to work with electron neutrino beams, see \cite{Kitano:2024kdv} for a recent proposal along these lines.

To overcome this problem, in Ref.~\cite{Schwetz:2021cuj} a method has been proposed, based on the well-known observation, that the transformation in \cref{eq:T} is formally equivalent to the transformation $L\to -L$:
\begin{equation}\label{eq:TL}
\mathcal{T}[P_{\alpha\to\beta}(L)] = P_{\alpha\to\beta}(-L) \,.  
\end{equation}
Hence, we can search for T violation by looking for an $L$-odd component of $P_{\nu_\alpha\to\nu_\beta}(L)$ considered as a function of $L$ at a fixed neutrino energy. It has been shown in \cite{Schwetz:2021cuj,Schwetz:2021thj} that under rather weak assumptions about neutrino properties such a test can be performed in a model-independent way in principle by combining measurements at three different baselines plus a near detector. 

In the present work we elaborate further on this idea, and we show that under certain conditions a test for T violation can be constructed by combining \textit{only two} experiments. We identify a largely model-independent observable $X_T$, built out of the observed probabilities $P_{\nu_\mu\to\nu_e}(L)$ at two baselines $L_1, L_2$ and at a near detector (ND) at $L\approx 0$, all determined at the same neutrino energy $E_\nu$, being defined as
\begin{equation}\label{eq:XT1}
  X_T \equiv P_{\nu_\mu\to\nu_e}(L_2) - P_{\nu_\mu\to\nu_e}(L_1) - \delta_0
  P_{\nu_\mu\to\nu_e}^{\rm ND} \,,
\end{equation}
where $\delta_0$ is a calculable coefficient. An analogous quantity can be derived also from the corresponding anti-neutrino measurements. The purpose of the ND measurement is to constrain zero-distance effects due to unitarity violation; it is not needed if 3-flavour unitarity is imposed as model assumption. As we show below, there exist combinations of $L_1, L_2$ and $E_\nu$, where $X_T$ is strictly positive if T is conserved, under modest assumptions on the underlying model of neutrino properties, similar to the ones adopted in \cite{Schwetz:2021cuj,Schwetz:2021thj}. Hence, if observations can establish that the combination $X_T$ is negative for the suitable combination of $L_1,L_2,E_\nu$, the time reversal symmetry is violated in nature.

Below we perform a systematic scan of possible $L_1,L_2,E_\nu$ combinations.
Indeed it turns out, that this test can be performed by combining the appearance probabilities measured at the T2HK~\cite{Hyper-Kamiokande:2018ofw} and the DUNE~\cite{DUNE:2020jqi,Abi:2020wmh} experiments at a neutrino energy around 0.86~GeV. In this work we will study this possibility in detail, and identify the experimental requirements for the test to work by performing simulations using the GLoBES software \cite{Huber:2004ka,Huber:2007ji}. While these requirements may turn out to be challenging, our observation opens an exciting opportunity for a model-independent search for T-violation with experiments already in preparation.

The outline of the paper is as follows. In \cref{sec:test} we introduce the general framework for the model independent test for T violation and then introduce the test for two experiments. We study in some detail the experimental configurations where the test can be applied and identify the T2HK/DUNE as suitable combination. In \cref{sec:simulation} we discuss the experimental  set up and analysis details used in our simulation of the T2HK and DUNE experiments and provide sensitivity estimates based on the expected event numbers and statistical errors. In 
\Cref{sec:results} we present our main numerical results and study in detail how the sensitivity depends on the various assumptions adopted in our analysis. In \cref{sec:conclusion} we summarize our findings. In \cref{sec:QFT} we elucidate the equivalence of time reversal symmetry and the transformation $L\to -L$ using the quantum field theory (QFT) framework for neutrino oscillations. In \cref{app:zero-distance} we provide details on the constraint related to the near-detector measurements at zero distance.

\section{Model-independent test for T violation}
\label{sec:test}

\subsection{Framework and assumptions}
\label{sec:framework}

Let us review the assumptions adopted for the T-violation test introduced in~\cite{Schwetz:2021cuj,Schwetz:2021thj} and discuss some minor modifications in the present work. The assumptions are:
\begin{enumerate}
\item[($i$)] We assume that the propagation of the three Standard Model (SM) neutrinos is governed by a hermitian Hamiltonian $H$, which depends on neutrino energy and the matter composition along the neutrino path. The evolution of the flavour state $|\psi\rangle$ is described by the Schr\"odinger equation
\begin{equation}\label{eq:schroedinger}
  i\partial_t|\psi\rangle = H(E_\nu) |\psi\rangle \,.
\end{equation}
\item[($ii$)] We are interested in experiments, where the matter density along the neutrino path can be taken as approximately constant and approximately the same for all experiments. This implies that the Hamiltonian is constant in space and time. The validity of this assumption and the size of corrections due to small deviations from it for the relevant experiments have been studied in \cite{Schwetz:2021thj}. 
\item[($iii$)] Let us denote flavour states relevant for detection ($d$) and production at the source ($s$) by $|\nu_\alpha^{d,s}\rangle$, respectively, and the eigenstates of the propagation Hamiltonian $H(E_\nu)$ by $|\nu_i\rangle$. Then we allow for arbitrary (non-unitary) mixing between them
\begin{equation}\label{eq:mixing}
  |\nu_\alpha^{s,d}\rangle = \sum_{i=1}^3 (N_{\alpha i}^{s,d})^* |\nu_i\rangle \qquad (\alpha = e,\mu,\tau)\,,
\end{equation}
where $^*$ denotes complex conjugation. We impose no a-priori constraints on the mixing parameters $N_{\alpha i}^{s,d}$. 
\item[($iv$)] We assume that effects of beyond SM physics for the Hamiltonian $H(E_\nu)$ of the evolution equation \cref{eq:schroedinger} are small, and the eigenvalues of the Hamiltonian and their energy dependence resembles approximately the one following from the effective neutrino mass squared differences in matter in the SM. We will quantify this requirement below.
\end{enumerate}
These assumptions cover of course the standard three-flavour oscillation framework and include also a broad range of beyond SM effects, such as non-standard neutrino interactions in production, propagation and detection as well as non-unitarity mixing. It allows also the presence of sterile neutrinos, as long as their mass-squared differences are much smaller or much larger than the two standard three-flavour mass-squared differences $\Delta m^2_{21} \approx 7.4\times 10^{-5}$~eV$^2$ and
 $\Delta m^2_{31} \approx 2.5\times 10^{-3}$~eV$^2$~\cite{Esteban:2020cvm} (Nu-Fit~5.3).
But assumption ($i$) excludes the possibility of neutrino decay on length-scales relevant for the experiments (but our framework does include neutrino decay with a decay-length much shorter than the distance between neutrino source and the closest near detector).

\bigskip

\textbf{Comment on time reversal.} The fundamental quantum mechanical evolution equation, \cref{eq:schroedinger}, describes evolution in time. Applied to neutrino oscillations, the common assumption is $t\approx x$, motivated by a neutrino wave packet picture with wave packets propagating close to the speed of light, and to consider evolution in space instead of time. For neutrino production at the position $x_s$ at time $t_s$ and detection at position $x_d$ and time $t_d$ the assumption $x\approx t$ implies $T \equiv t_d - t_s = x_d-x_s \equiv L$. Hence, the time reversal transformation $t\to -t$ leads to an effective transformation in space, $L\to -L$, see also the discussion in Ref.~\cite{Schwetz:2021thj} in the context of non-standard mixing scenarios. While the equivalence of the transformations \cref{eq:T} and \cref{eq:TL} follows immediately from the standard formula for oscillation probabilities, this somewhat hand-waving argumentation relies on the association $x\approx t$, which emerges as an a-posteriori result of a consistent wave-packet treatment~\cite{Akhmedov:2010ms}. We show in \cref{sec:QFT}, that \cref{eq:T,eq:TL} and their equivalence can be derived also from a quantum-field theoretical approach to neutrino oscillations \cite{Beuthe:2001rc}.

\bigskip

\textbf{Transition probabilities.} For the sake of definiteness, let us consider the $\nu_\mu\to\nu_e$ appearance probability, introducing the abbreviation $P \equiv P_{\nu_\mu\to\nu_e}$, with the $L$ and $E_\nu$ dependence left implicit. All arguments apply in a straight-forward way to anti-neutrinos as well. Under the above stated assumptions the probability is given by
\begin{equation}\label{eq:P}
  P = \left| \sum_{i=1}^3 c_i e^{-i\lambda_iL}\right|^2 \,,
  \qquad c_i \equiv N_{\mu i}^{s*} N_{e i}^d \,,
\end{equation}
where $\lambda_i$ are the eigenvalues of the effective Hamiltonian from \cref{eq:schroedinger}. We write $P$ as
\begin{align}
  P 
    &= \left|  c_2 (e^{-i(\lambda_2 - \lambda_1)L} - 1)+ c_3 (e^{-i(\lambda_3 - \lambda_1)L} - 1) + \epsilon \right|^2 \,,   
\end{align}
where we have defined the parameter
\begin{align}\label{eq:epsilon}
  \epsilon \equiv \sum_{i=1}^3 c_i \,.
\end{align}
In the unitary-mixing case, $N_{\alpha i}^s = N_{\alpha i}^d = U_{\alpha i}$, and it follows from the definition of $c_i$ in \cref{eq:P} that $\epsilon = 0$ for unitary mixing. Hence, $\epsilon$ describes deviation from unitarity and leads to a ``zero-distance effect'', which we generically denote by ``near detector'' (ND):
\begin{equation}\label{eq:ND}
  P^{\rm ND} \equiv P(L\to 0) = |\epsilon|^2 \,.
\end{equation}
For the analytical discussion it is convenient to use the three independent parameters $\epsilon, c_2,c_3$ instead of $c_1,c_2,c_3$.
One can split the probability into T-even and T-odd parts:
\begin{align}
  P &= P_{\rm even} + P_{\rm odd} \,,   
\end{align}
with
\begin{align}
  P_{\rm even} &= |\epsilon|^2
  + 4{\rm Re}[c_2^*(c_2-\epsilon)]\sin^2\phi_{21}
  + 4{\rm Re}[c_3^*(c_3-\epsilon)]\sin^2\phi_{31} \nonumber\\
  &+ 8{\rm Re}[c_2^*c_3]\sin\phi_{21}\sin\phi_{31}\cos(\phi_{31}-\phi_{21}) \\[2mm]
  P_{\rm odd} &= 2{\rm Im}[\epsilon^*c_2] \sin 2\phi_{21} + 2{\rm Im}[\epsilon^*c_3] \sin 2\phi_{31}
  \nonumber\\
  &+ 8{\rm Im}[c_2^*c_3]\sin\phi_{21}\sin\phi_{31}\sin(\phi_{31}-\phi_{21}) \,,
\end{align}
where
\begin{align}
  \phi_{ij} \equiv \frac{\lambda_i-\lambda_j}{2}L \,.
\end{align} 

As stated in assumption ($iv$) above, we assume that new-physics contributions to $\lambda_i$ are small, i.e., 
\begin{align}\label{eq:phi}
  \phi_{ij} \approx \frac{\Delta m^2_{ij,{\rm eff}}(E_\nu) L}{4 E_\nu} \,,
\end{align}
where $\Delta m^2_{ij,{\rm eff}}(E_\nu)$ are the effective mass-squared differences in matter, assuming the standard matter effect. In our numerical work we obtain them by diagonalizing the effective Hamiltonian in matter numerically, assuming the best fit oscillation parameters from~\cite{Esteban:2020cvm} (Nu-Fit~5.3) and an average matter density of $\rho = 2.84\,\rm g\, cm^{-3}$. 

\subsection{T-violation test for two experiments}
\label{sec:2exps}

Let us consider the T-even part, i.e., we assume that T is conserved by the fundamental theory. This implies that the parameters $\epsilon, c_2,c_3$ are real. We write
\begin{align}\label{eq:Teven}
  P_{\rm even} =  \gamma_2 c_2(c_2-\epsilon) + \gamma_3 c_3(c_3-\epsilon) +
  \gamma_{23} c_2c_3  + \epsilon^2
\end{align}
with the abbreviations
\begin{equation}\label{eq:gamma}
  \begin{split}
  \gamma_i &= 4\sin^2\phi_{i1} \quad (i=2,3) \,,\\
  \gamma_{23} &= 8 \sin\phi_{21}\sin\phi_{31}\cos(\phi_{31}-\phi_{21}) \,. 
  \end{split}
\end{equation}  
These coefficients are functions of neutrino energy and baseline.
Following~\cite{Schwetz:2021cuj}, we can establish that the fundamental theory violates T
if data cannot be fitted with the T-even part alone. 
Since \cref{eq:Teven} depends on three parameters ($c_2, c_3, \epsilon$) one may conclude that there is always a fit for two experiments plus a near detector, which provide three data points. In the following we show, that this argument is actually not true and under certain conditions the quadratic nature of the parameter dependence does not provide a solution for three data points, even imposing no further condition on $c_2, c_3, \epsilon$.

Let us consider the difference of the appearance probability at two baselines, $L_1$ and $L_2$ but at the same energy:
\begin{align}
  P_{\rm even}(L_2) - P_{\rm even}(L_1) 
  =  \delta_2 c_2(c_2-\epsilon) + \delta_3 c_3(c_3-\epsilon) +
  \delta_{23} c_2c_3 
\end{align}
with 
\begin{align}\label{eq:delta}
  \delta_i = \gamma_i(L_2) - \gamma_i(L_1) \quad (i=2,3,23) \,.
\end{align}
If $\epsilon \neq 0$, a suitable shift of variables can be performed
\begin{equation}
 c_2 \to c_2 + \epsilon \frac{\delta_3\delta_{23} - 2\delta_2\delta_3}{\delta_{23}^2 - 4\delta_2\delta_3} \,,\qquad 
 c_3 \to c_3 + \epsilon \frac{\delta_2\delta_{23} - 2\delta_2\delta_3}{\delta_{23}^2 - 4\delta_2\delta_3} \,,
\end{equation}
such that 
\begin{align} \label{eq:XT}
  X_T \equiv P_{\rm even}(L_2) - P_{\rm even}(L_1) - \epsilon^2\delta_0
  =  \delta_2 c_2^2 + \delta_3 c_3^2 + \delta_{23} c_2c_3  
\end{align}
with
\begin{align}\label{eq:delta0}
  \delta_0 &= \frac{\delta_2+\delta_3 - \delta_{23}}{\delta_{23}^2/(\delta_2\delta_3) - 4} \,.
\end{align}
In \cref{eq:XT} we have defined the quantity $X_T$ introduced already in \cref{eq:XT1}.

Without loss of generality we assume $\delta_2>0$, which can be achieved by ordering $L_1$ and $L_2$ accordingly.\footnote{Numerically we have $\phi_{21} < \pi/2$ for the experiments of interest, which implies that $\delta_2 > 0$ for $L_2 > L_1$.} The important observation is now that the right-hand side of \cref{eq:XT} is a \textit{non-negative function} of $c_2$ and $c_3$ iff 
\begin{align}
  & \delta_3>0 \quad \text{and}\quad \delta_2>0 \,, \quad\text{and} \label{eq:cond1}\\
  & |\delta_{23}| < 2 \sqrt{\delta_2\delta_3} \,. \label{eq:cond2}
\end{align}

Using \cref{eq:ND}, we can now consider the observed value of the quantity $X_T$,
\begin{equation}\label{eq:XTobs}
  X_T^{\rm obs} = P_{\nu_\mu\to\nu_e}^{\rm obs}(L_2) - P_{\nu_\mu\to\nu_e}^{\rm obs}(L_1)
- \delta_0 P_{\nu_\mu\to\nu_e}^{\rm ND,obs} \,,
\end{equation}
and obtain the following test for T violation:

\bigskip

\framebox{\parbox{0.9\textwidth}{
If it can be established within experimental uncertainties that $X_T^{\rm obs} < 0$  
and the conditions \cref{eq:cond1,eq:cond2} are fulfilled then T has to be violated in Nature.
}}

\bigskip

In other words, if $X_T^{\rm obs} < 0$  and  \cref{eq:cond1,eq:cond2} hold, a T-odd component needs to be present in the transition probability to make $X_T$ negative, because \cref{eq:XT} implies that the T-even contributions alone have to be non-negative, regardless of the values of the mixing parameters $c_i$. 

Note that the condition $X_T < 0$ is somewhat conservative. We can still use 
$P^{\rm ND,obs}$ to fix $\epsilon$ and $P^{\rm obs}$ at one of the far positions, let's say $L_1$, to impose a constraint on $c_2$ and $c_3$. If $P^{\rm obs}(L_1) \neq P^{\rm ND,obs}$ then $c_2$ and $c_3$ cannot both be zero, which implies that the right-hand side of \cref{eq:XT} is positive. Numerically, however, we find that if no further restriction on $c_2$ and $c_3$ is imposed, there is always a combination of them which makes the right-hand side of \cref{eq:XT} very small while keeping $P_{\rm even}(L_1) = P^{\rm obs}(L_{1})$. In our numerical analysis presented in the next section these effects are consistently taken into account, including also experimental uncertainties.

\begin{figure}
  \centering
  \includegraphics[width=0.9\textwidth]{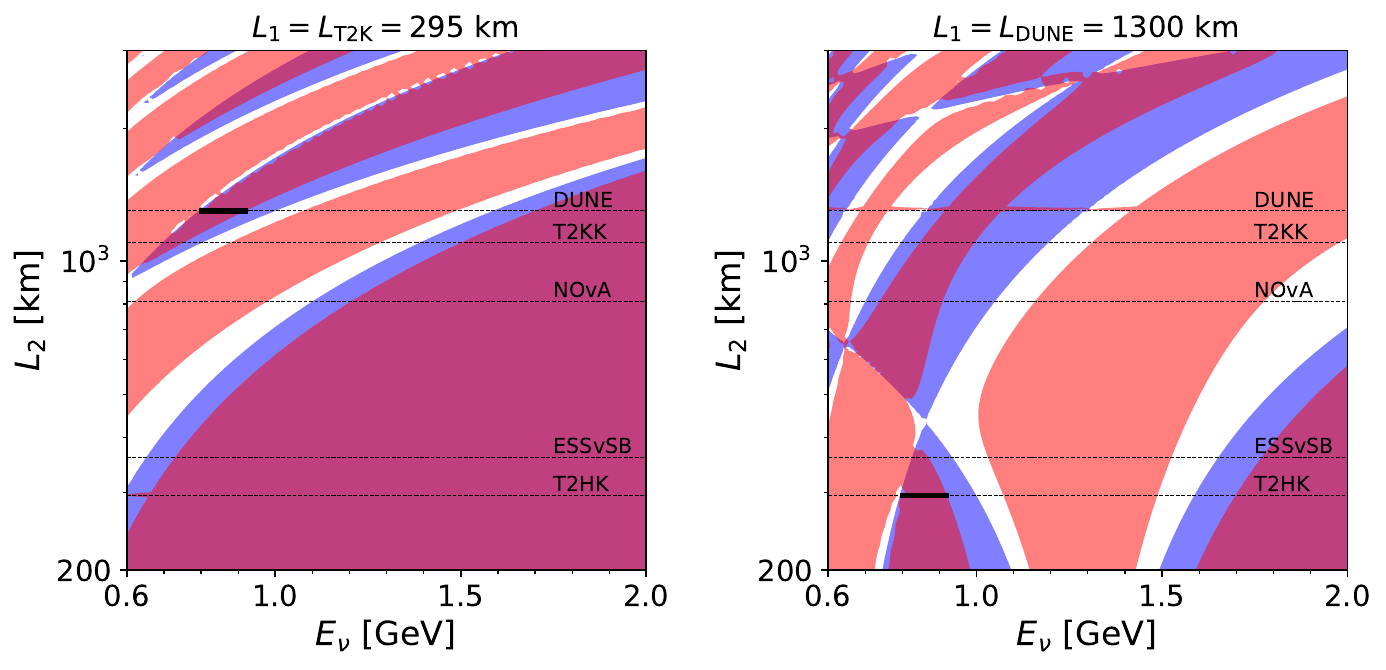}\\[3mm]
  \includegraphics[width=0.9\textwidth]{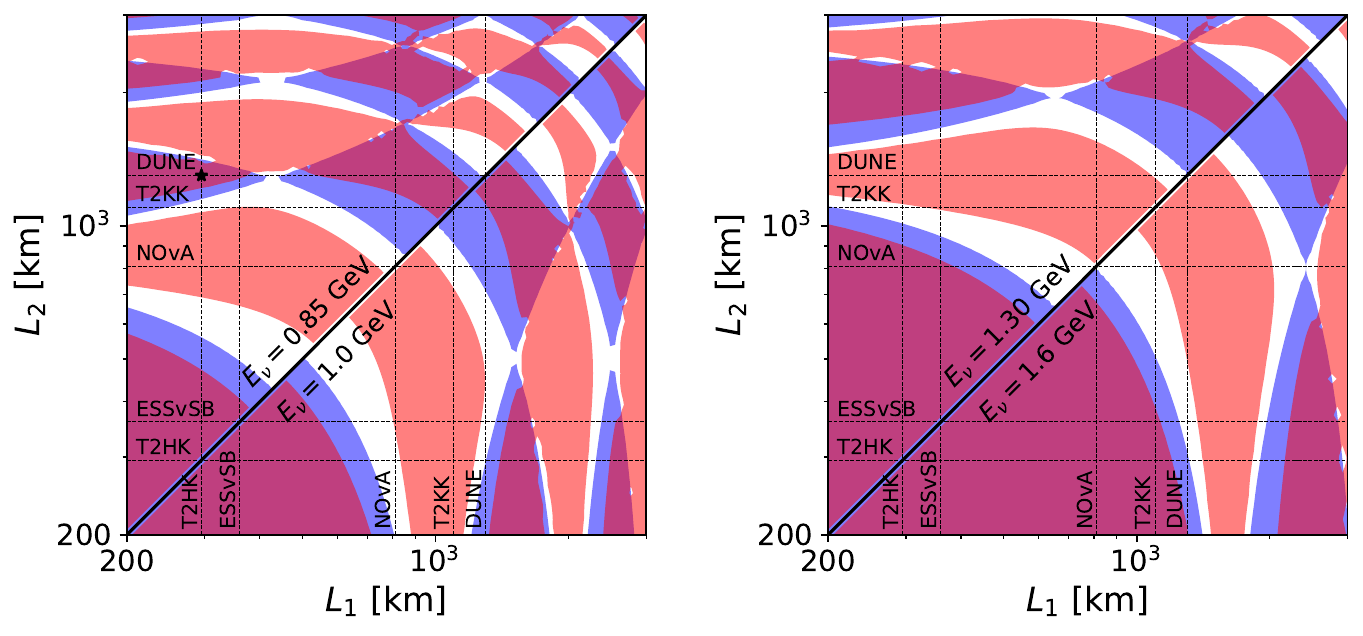}
  \mycaption{Regions in energy and distance where the conditions \cref{eq:cond1,eq:cond2} are fulfilled. Blue and red regions correspond to \cref{eq:cond1} and \cref{eq:cond2}, respectively, and purple regions to both conditions simultaneously. In the upper panels we fix $L_1$ to $L_{\rm T2K}$ and $L_{\rm DUNE}$, respectively, vary $L_2$ on the vertical axis and show the neutrino energy on the horizontal axis. In the lower panels we show the two distances $L_{1,2}$ on the axes for four fixed energies $E_\nu = 0.85, 1.0, 1.3, 1.6$~GeV in each triangle section of the panels, respectively. We assume neutrinos and normal mass ordering.}
  \label{fig:condition}
\end{figure}

\bigskip

\textbf{Can the conditions be fulfilled in realistic experiments?} Note that the coefficients $\delta_i$ are functions of neutrino energy and baseline and depend on the effective energy eigenvalues in matter, see \cref{eq:delta,eq:gamma,eq:phi}. Hence, according to assumption ($iv$) they are primarily determined by the neutrino mass-squared differences and show a (typically weak) dependence on the leptonic mixing angles due to the matter effect. We adopt the 
best-fit values from~\cite{Esteban:2020cvm} (Nu-Fit~5.3) to calculate $\delta_i$ as a function of $L_1,L_2$ and $E_\nu$ and search for regions where the conditions  \cref{eq:cond1,eq:cond2} are satisfied. The results of this analysis are shown in \cref{fig:condition}, where in the purple regions both conditions are satisfied simultaneously. To guide the eye, we indicate by the dotted lines the baselines of a few representative future long-baseline experiments, namely
T2HK~\cite{Hyper-Kamiokande:2018ofw} ($L=295$~km),
ESS$\nu$SB~\cite{Alekou:2022emd} ($L=360$~km),   
T2KK~\cite{Hyper-Kamiokande:2016srs} ($L=1100$~km), and
DUNE~\cite{DUNE:2020jqi,Abi:2020wmh} ($L=1300$~km). Several comments are in order:
\begin{enumerate}
\item 
Purple regions of small $L$ and/or large $E_\nu$ (bottom-right corners in upper pannels and bottom-left corners in lower pannels) correspond to regions of very low values of the probabilities below the first oscillation maxima; here the conditions are formally satisfied, but irrelevant for practical purposes due to negligible event numbers.
\item
  Focusing on the upper left panel with $L_1 = 295$~km corresponding to the T2HK baseline, we find a potentially sensitive region at $L_2$ corresponding to the DUNE baseline around $E_\nu \approx 0.86$~GeV. This is also visible in the upper-right and lower-left panels. This energy is well suited for the T2HK beam, but at the low-energy tail for DUNE. In the next section we will study in detail the experimental requirements to explore that region.
  \item
  Apart from the T2HK/DUNE region we find also a potential region for T2HK/T2KK around 
$E_\nu \approx 0.75$~GeV. Numerically it turns out that this combination does not provide as good sensitivity as the T2HK/DUNE combination. Otherwise, none of the other baselines crosses a suitable purple region when combined with T2HK.
\item
  In the upper-right panel, with $L_1 = 1300$~km equal to the DUNE baseline, we find in addition to T2HK a potential combination with NOvA at $E_\nu \approx 0.9$~GeV (also visible in the lower-left panel). However, at these energies, NOvA has no events and therefore this window cannot be used. Furthermore, the conditions are met as well for the T2KK/DUNE combination around $E_\nu \approx 0.8$~GeV. In this case the baselines $L_1$ and $L_2$ are comparable, and the probability difference is very small, which again prevents us to use this combination in practice.
\item
  Apart from these cases, no other combination of the selected experiments falls in the region where all conditions for the $X_T$ test can be fulfilled for useful energies.
  For the figure we assumed neutrino mode and the normal neutrino mass ordering. For other combinations, the regions shift slightly due to the modified matter effect, but the qualitative picture remains the same. Specifically, the energy intervals where the conditions \cref{eq:cond1,eq:cond2} are fulfilled for the T2HK/DUNE combination are:
\begin{equation}\label{eq:bin} 
  \begin{split}
    E_\nu \in [0.80,0.92] \,{\rm GeV}\qquad \text{(neutrinos/NO and anti-neutrinos/IO)} \,, \\
    E_\nu \in [0.86,0.99] \,{\rm GeV}\qquad \text{(neutrinos/IO and anti-neutrinos/NO)} \,.
  \end{split}
\end{equation}
\item In addition to the conditions \cref{eq:cond1,eq:cond2}, also $X_T^{\rm obs}$ needs to be negative for this $L_1,L_2,E_\nu$ combination in order to establish T violation. This will depend on the actual mechanism for T violation realised in nature. The simplest hypothesis is just standard 3-flavour oscillations, with T violation induced by the Dirac phase in the PMNS matrix, $\delta_{\rm CP}$. We denote this case by ``Standard Model'' (SM) in the following. Indeed, for the T2HK/DUNE combination at $E_\nu \approx 0.86$~GeV we find that $X_T^{\rm SM}(\delta_{\rm CP} \approx \pi/2) \approx -0.012$ for neutrinos, while it is positive for $\delta_{\rm CP} \approx 3\pi/2$. For anti-neutrinos the situation is the opposite: $X_T^{\rm SM}$ is positive (negative) for $\delta_{\rm CP} \approx \pi/2 \,(3\pi/2)$. For these estimates we assumed $P^{\rm ND} \approx 0$. The coefficient relevant for $P^{\rm ND}$ is $\delta_0 \approx -0.24$. These numbers set the required precision on the three probabilities to establish $X_T^{\rm obs} < 0$ at a useful significance (see detailed simulations below).
\item
Finally, going beyond the considered experiment proposals, we see from \cref{fig:condition} that there is also a potentially interesting region for combining a hypothetical experiment at $L_2 \simeq 2000-3000$~km with DUNE at energies between 1 and 1.6~GeV (see right panels). In that region it turns out that $X_T^{\rm SM}$ is positive for neutrinos (for both, $\delta_{\rm CP} = 90^\circ$ and $270^\circ$). But the test could potentially work for anti-neutrinos in the standard 3-flavour case, where $X_T^{\rm SM}$ can become negative.
\end{enumerate}

In summary, the conditions \cref{eq:cond1,eq:cond2} are necessary, but not sufficient that the $X_T$ test can be applied realistically. We find that the T2HK/DUNE combination is the most promising configuration, which we therefore have investigated in some detail with numerical simulations.

\section{Simulation details} 
\label{sec:simulation}

In \cref{sec:experimental-setup} we give the experimental specifications of the DUNE and T2HK setups assumed in this work to explore the sensitivity to T-violation, which is then followed in \cref{sec:analysis} by a description of the analysis adopted for our statistical analysis. In \cref{sec:sens-estimate} we provide an analytical estimate of the sensitivity based on the expected event numbers.

\subsection{Experimental setups}
\label{sec:experimental-setup}

{\bf DUNE} is an upcoming long-baseline accelerator neutrino experiment designed to explore the nature of neutrinos by sending them from Fermilab to a far-detector situated deep underground at the Sanford Lab in South Dakota at a  distance of 1300~km. To estimate the experiment's sensitivity, we utilize the configurations outlined in the Technical Design Report (TDR) \cite{DUNE:2020jqi,DUNE:2021cuw}, along with an alternative configuration featuring enhanced energy resolution as suggested in Refs.~\cite{DeRomeri:2016qwo,Friedland:2018vry} (see also \cite{Kopp:2024lch,Chatterjee:2021wac}).
The TDR configuration includes a 40~kt Liquid Argon Time Projection Chamber (LArTPC) as the far detector (current planning considers a staged approach with up to 4 detector modules of 17~kt each), paired with a 120~GeV proton beam delivering 1.2~MW of beam power, which translates to $1.1 \times 10^{21}$ protons on target (P.O.T.) per year.
For further details on systematic errors and efficiencies, please refer to Ref.~\cite{DUNE:2021cuw}.
\\

\noindent {\bf T2HK} (Tokai to Hyper-Kamiokande) is an off-axis, accelerator-based future superbeam experiment with a 295 km baseline. To estimate the detector's physics potential, we adhere to the experimental configurations outlined in~\cite{Hyper-Kamiokande:2018ofw}. This experiment will utilize the same 30~GeV proton beam from the J-PARC facility, previously used for T2K, to generate (anti)neutrino fluxes. The Water Cherenkov far detector is expected to have a fiducial volume of 187~kt, and the total exposure will be $1.3 \,\rm{MW}\times 10\times 10^7$ seconds, equivalent to $2.7\times 10^{22}$ protons on target (P.O.T.).
In this simplified scenario, we consider an uncorrelated 5\% (3.5\%) signal normalization error, a 10\% background normalization error, and a 5\% energy calibration error for both $\nu$ and $\bar{\nu}$ appearance (disappearance) channels.\\

\begin{figure}[t]
    \includegraphics[width=0.5\textwidth]{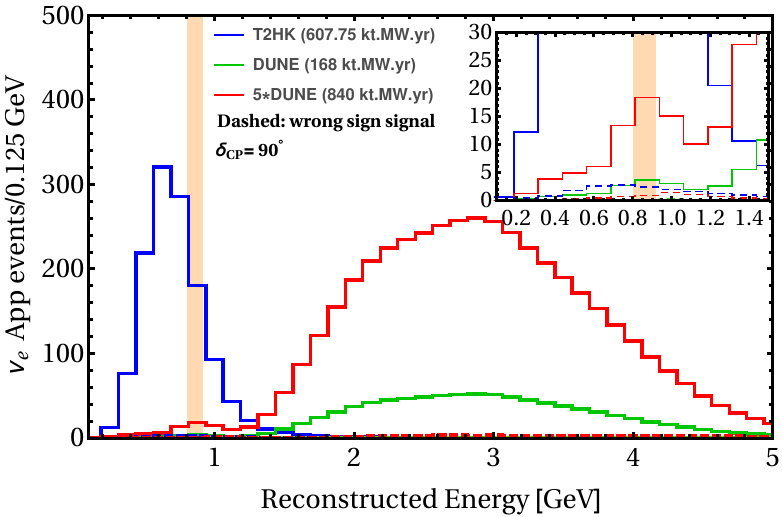}
    \includegraphics[width=0.5\textwidth]{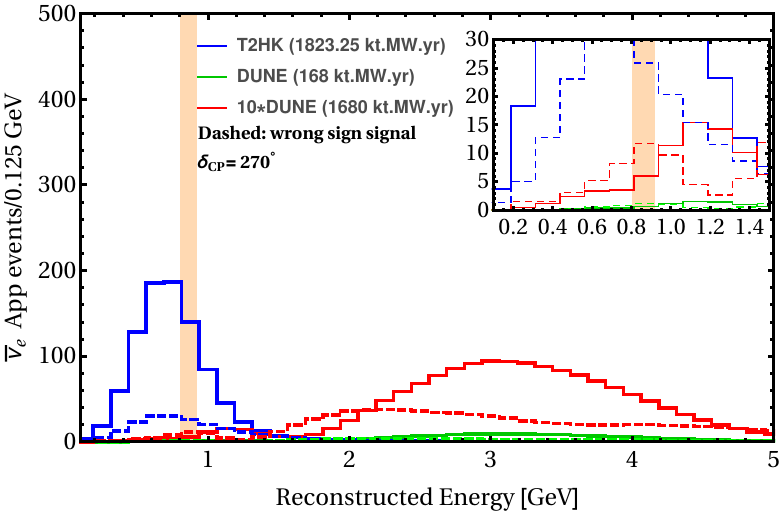}
        \mycaption{Number of $\nu_\mu\to\nu_e$ (left) and $\bar\nu_\mu\to\bar\nu_e$ (right) signal events per 0.125~GeV reconstructed neutrino energy bins. For T2HK we assume an exposure of 607.75 (1823.25)~kt~MW~yr for neutrino (antineutrino) running. For DUNE we show spectra for a nominal exposure of 168~kt~MW~yr by green curves, as well as exposures increased by a factor 5 (10) for neutrinos (antineutrinos) as red curves. Dashed curves indicate events due to the wrong-sign beam component, i.e., $\bar\nu_\mu\to\bar\nu_e$ for the left panel (hardly visible) and $\nu_\mu\to\nu_e$ for the right panel. The vertical bar indicates the energy bin sensitive to the T-violation test. The insets show a zoom into the relevant energy range. 
        We assume standard oscillations with the parameters given in \cref{tab:osc-params} and $\delta_{\rm CP} = 90^\circ$ (left panel) and $\delta_{\rm CP} = 270^\circ$ (right panel).
      }
    \label{fig:event-spectra}
\end{figure}

\Cref{fig:event-spectra} shows the expected signal spectra for the $\nu_\mu\to\nu_e$ appearance channel, assuming standard three-flavour oscillations with parameters given below in \cref{tab:osc-params} and $\delta_{\rm CP} = 90^\circ\,(270^\circ)$ for neutrino (antineutrino) beam mode. In the left panel we consider the neutrino mode, where for T2HK we have assumed an exposure of 608~kt~MW~yr, which corresponds to about 2.5~yr of neutrino beam running with the above mentioned assumptions on detector mass and beam power. We see from the figure, that the sensitive energy window identified in \cref{eq:bin} falls close to the peak of the event spectrum in T2HK with about 180 events. For DUNE, however, the relevant energy range is located at the low-energy tail of the event spectrum, suffering from low event numbers for a nominal exposure of 168~kt~MW~yr.
Therefore, we have assumed an optimistic exposure of 840~kt~MW~yr, which would give about 18 events in the sensitive energy bin for the chosen oscillation parameters. According to current detector installation and beam power planning \cite{IRN-talk}, this exposure will be achieved roughly after 13~years of operation. Note that here we assumed that the total exposure is collected with neutrino running. Unless otherwise specified, this will be our default exposure assumptions for both T2HK and DUNE in the following.

In the right panel we show event spectra for antineutrino running. With the T2HK exposure of 1823~kt~MW~yr we find $\simeq 140$ events in the relevant energy range, however, event numbers for DUNE are very small. Even an (unrealistically) high exposure of 1680~kt~MW~yr pure antineutrino exposure would lead to only 6 events in the relevant energy range and a ``wrong-sign'' beam component giving an even larger signal of about 12 events. Hence, we expect that the application of our proposed test for antineutrinos will suffer from too low statistics.

Finally, we see from \cref{fig:event-spectra} that the sensitive energy bin is located at the 1st and 2nd oscillation maxima for the T2HK and DUNE baselines, respectively. Hence, the T-violation test is based on the comparison of the probabilities at the 1st and 2nd oscillation maxima, see also the discussion in \cite{Schwetz:2021cuj} (supplementary material).

\subsection{Analysis details} 
\label{sec:analysis}

\begin{table}[t]
  \centering
  \begin{tabular}{c@{\quad}c@{\quad}c@{\quad}c@{\quad}c}
    \hline\hline
    $\sin^2{\theta_{12}}$ & $\sin^2{\theta_{13}}$ & $\sin^2{\theta_{23}}$ & $\Delta m^2_{21}$ [eV$^2$]& $|\Delta m^2_{3\ell}|$  [eV$^2$] \\
    0.307 & 0.022 & 0.572 & $7.41\times 10^{-5}$ & $2.51\times 10^{-3}$ \\
    \hline\hline
  \end{tabular}
  \mycaption{Standard three-flavour parameters adopted in the numerical analysis, with $\ell = 1$ (2) for normal (inverted) mass ordering~\cite{Esteban:2020cvm} (NuFit 5.3). }
\label{tab:osc-params}
\end{table}

To estimate the sensitivity to T-violation, we have used the GLoBES~\cite{Huber:2004ka,Huber:2007ji} package with the required modifications of the probability engine. We have explicitly implemented the transition probabilities according to
\cref{eq:P}. In our analysis we only use the appearance channel. We assume that the eigenvalues of the Hamiltonian are determined according to \cref{eq:phi} by the effective mass-squared differences in matter as in the SM, where the values of $\Delta m^2_{21}$ and $\Delta m^2_{31}$ are fixed due to external constraints with sufficient precision (to be quantified below). Hence, the free parameters in our fit are the coefficients $c_i$ ($i=1,2,3$) introduced in \cref{eq:P}.

The test for T violation is based on the comparison of transition probabilities at fixed energies. Therefore, we choose identical bins in reconstructed neutrino energy for both, T2HK and DUNE, and
consider a $\chi^2$-function for a given energy bin:
\begin{equation}\label{eq:chi2k}
  \chi^2_k(c_i) = \sum_x  \min\limits_{\xi_x}\left[
    2\left(N_k^x(c_i,\xi_x) -N_k^{x,\rm obs} - N_k^{x,\rm obs} \ln \frac{N_k^x(c_i,\xi_x)}{N_k^{x,\rm obs}}\right) + \sum_{\xi_x} \frac{\xi_x^2}{\sigma_\xi^2} \right] 
    + \left[\frac{\epsilon(c_i)^2}{\sigma_\epsilon}\right]^2
    \,.
\end{equation}
Here $k$ labels the energy bin and $N_k^x(c_i,\xi_x)$ is the number of events predicted in the model \cref{eq:P} for experiment $x={\rm T2HK, DUNE}$, calculated including backgrounds and various systematics as described in \cref{sec:experimental-setup}. The latter are parametrised by pull parameters, generically denoted by $\xi_x$ in \cref{eq:chi2k}.
$N_k^{x,\rm obs}$ is the corresponding ``observed'' number of events, which will depend on the true mechanism of neutrino conversion realised in Nature. In this study we will---as a specific example---always assume the standard three-flavour scenario and calculate $N_k^{x,\rm obs}$ accordingly, using the oscillation parameters shown in \cref{tab:osc-params}.
Then we study the sensitivity of the T-violation test as a function
of the assumed true value of $\delta_{\rm CP}$, as $N_k^{x,\rm obs}(\delta_{\rm CP})$. Hence, 
$\chi^2_k(c_i) \to \chi^2_k(c_i,\delta_{\rm CP}^{\rm true})$.
Obviously, there can only be sensitivity for T-violation for values $\delta_{\rm CP} \neq 0,\pi$. To calculate 
$N_k^{x,\rm obs}$ in the standard three-flavour case we use a line-averaged constant matter density of 2.84~g/$\rm{cm}^3$~\cite{stacey:1977,PREM:1981} for both, DUNE and T2HK, which is a good approximation for these baselines \cite{Schwetz:2021thj}. The same value is then adopted to calculate the oscillation frequencies in the T conserving model according to \cref{eq:phi}.\footnote{In this approximation the model implemented by \cref{eq:P} can reproduce event numbers for T conservation $\delta_{\rm CP} = 0,\pi$ exactly.}

The last term in \cref{eq:chi2k} takes into account a constraint on the zero-distance effect, which is implemented as external prior in GLoBES. The parameter $\epsilon(c_i) = \sum_i c_i$ has been defined in \cref{eq:epsilon} and it induces non-zero transitions at zero-distance due to unitarity violation. We assume that its size is constrained with an effective uncertainty $\sigma_\epsilon$, which emerges from a combination of near-detector measurements, correlated systematic uncertainties as well as external constraints on non-unitarity, see \cref{app:zero-distance} for a detailed discussion. Recent global analyses on non-unitarity can be found  e.g., in 
Refs.~\cite{Ellis:2020hus,Hu:2020oba,Forero:2021azc,Blennow:2023mqx}. For instance, the parameter $\eta_{e\mu}$ used in Ref.~\cite{Blennow:2023mqx} is related to our $\epsilon$ by $|\epsilon| \approx 2|\eta_{e\mu}|$ and the results of Ref.~\cite{Blennow:2023mqx} imply an external constraint of $|\epsilon| < 1\, (1.4) \times 10^{-5}$ at 68\% (95\%)~CL (``GUV analysis''), which taken at face value would imply a negligible zero-distance transition probability $P_{\rm ND} = \epsilon^2$ for all practical purposes.    
We note, however, that typically these analyses are to some extent model dependent, for instance by assuming a certain energy dependence of non-unitarity effects. Here we allow the $c_i$ coefficients to vary independently in each energy bin, i.e., remaining fully agnostic about their energy dependence. The analysis in \cite{Blennow:2023mqx} assumes non-unitarity due to ``heavy'' new physics; in the presence of light sterile neutrinos non-unitarity effects may be larger; e.g., ref.~\cite{Forero:2021azc} finds $P_{\rm ND} < 4\times 10^{-4}$ (90\%~CL).
Furthermore, in the presence of correlated uncertainties (such as flux uncertainties) the effective uncertainty on $\epsilon^2$ will be much larger than the external constraint mentioned above and dominated by the accuracy of near detector measurements, see \cref{app:zero-distance}.
Below we will study how the sensitivity of our T violation test depends on the size of $\sigma_\epsilon$. If not stated otherwise, the default assumption is $\sigma_\epsilon = 10^{-3}$. 

Departing from \cref{eq:chi2k}, we define
\begin{equation}\label{eq:chi2T}
  \Delta \chi^2_T = \sum_k \min\limits_{c_i}\left[\chi^2_k(c_i)\right] \,,
\end{equation}
which we interpret as sensitivity to T violation by evaluating it for 1~dof. The statistical interpretation is as follows: for a single bin in the Gaussian approximation, $\sqrt{\Delta \chi^2_T}$ can be interpreted as the number of standard deviations at which the observable $X_T < X_T^{\rm min}$, where $X_T^{\rm min}$ is the minimum value of $X_T$ allowed for T conservation. 
(As proven in \cref{sec:2exps} we always have $X_T^{\rm min} \ge 0$.) When using several bins, the significances are just added by adding the individual $\chi^2$'s for each bin.

Note that the sum over energy bins $k$ is done \emph{after} minimising with respect to the parameters $c_i$ in \cref{eq:chi2T}, contrary to the usual model-dependent standard fitting procedure. Hence, in our model-independent approach we allow different best-fit parameters $c_i$ in each energy bin, and therefore the number of free parameters in the fit depends on the number of bins. This procedure is adopted to allow for an unknown energy dependence of possible new physics effects. As a result a subtle interplay between the chosen number of bins as well as the assumed energy resolution emerges. From the discussion in \cref{sec:2exps} we expect that only in the true neutrino energy interval \cref{eq:bin} there is sensitivity. However, because of smearing effects due to finite energy resolution also neighbouring bins will show some sensitivity. As a default configuration we will adopt 3 bins, where the central bin is given by the interval in \cref{eq:bin}, plus one bin of the same size above and below this interval. Below we will discuss also the dependence of the results on this choice.

\subsection{Sensitivity estimate}
\label{sec:sens-estimate}

Before we discuss the results of our statistical analysis, let us provide a rough estimate for the sensitivity to T violation of the considered T2HK and DUNE configurations. We consider the quantity $X_T$ from \cref{eq:XTobs} and estimate the significance with which it is negative assuming neutrino exposures and the benchmark parameters as adopted in \cref{fig:event-spectra}. For $E_\nu = 0.85$~GeV  we find
\begin{equation}
  P(L_{\rm DUNE}) = 0.0233\,,\quad
  P(L_{\rm T2HK}) = 0.0357\,,\quad
  P(L_{\rm ND}) = 0\quad \Rightarrow\quad 
  X_T = -0.0124 \,.
\end{equation}
We can estimate the statistical uncertainty on $X_T$ by using the number of events predicted in the sensitive energy bin of $N_{\rm DUNE} \approx 18$ and $N_{\rm T2HK} \approx 180$ (c.f.\ \cref{fig:event-spectra}) as
\begin{align}
  \sigma_{X_T} &= \sqrt{\frac{P(L_{\rm DUNE})^2}{N_{\rm DUNE}} + \frac{P(L_{\rm T2HK})^2}{N_{\rm T2HK}} +
  \delta_0^2 \sigma_\epsilon^2} \label{eq:stat-error1}\\
  &\approx 3.2\times 10^{-3}\left[ 3.0 \, \frac{18}{N_{\rm DUNE}} + 0.71 \, \frac{180}{N_{\rm T2HK}} +
    0.006 \left(\frac{\sigma_\epsilon}{0.001}\right)^2\right]^{1/2} \,, \label{eq:stat-error}
\end{align}
where $\sigma_\epsilon$ is the effective uncertainty on the zero-distance effect (see \cref{app:zero-distance} for a discussion) and $\delta_0 \approx -0.243$. Hence, we obtain a significance of $X_T$ being negative of $|X_T|/\sigma_{X_T} \approx 2$ standard deviations. This estimate turns out to be in rough agreement with the more elaborate statistical analysis presented below (which typically will lead to slightly better sensitivities, depending on the assumed energy resolution). 

\section{Results}\label{sec:results}

\subsection{Exposure and energy resolution}

\begin{figure}[t]
    \includegraphics[width=0.5\linewidth]{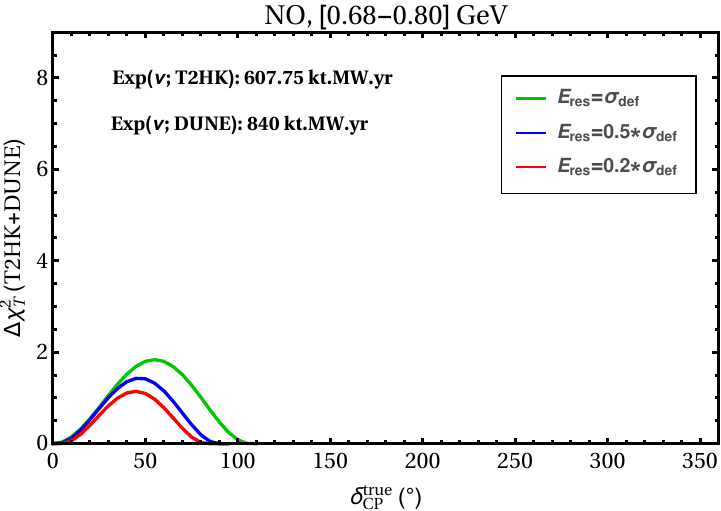}
    \includegraphics[width=0.5\linewidth]{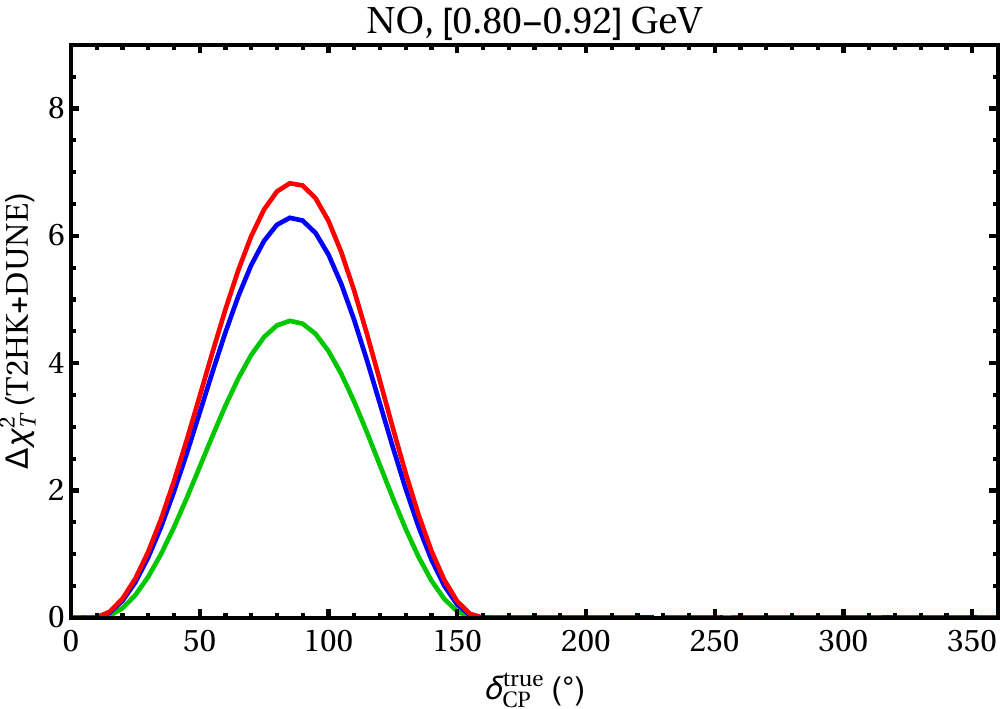}
    \includegraphics[width=0.5\linewidth]{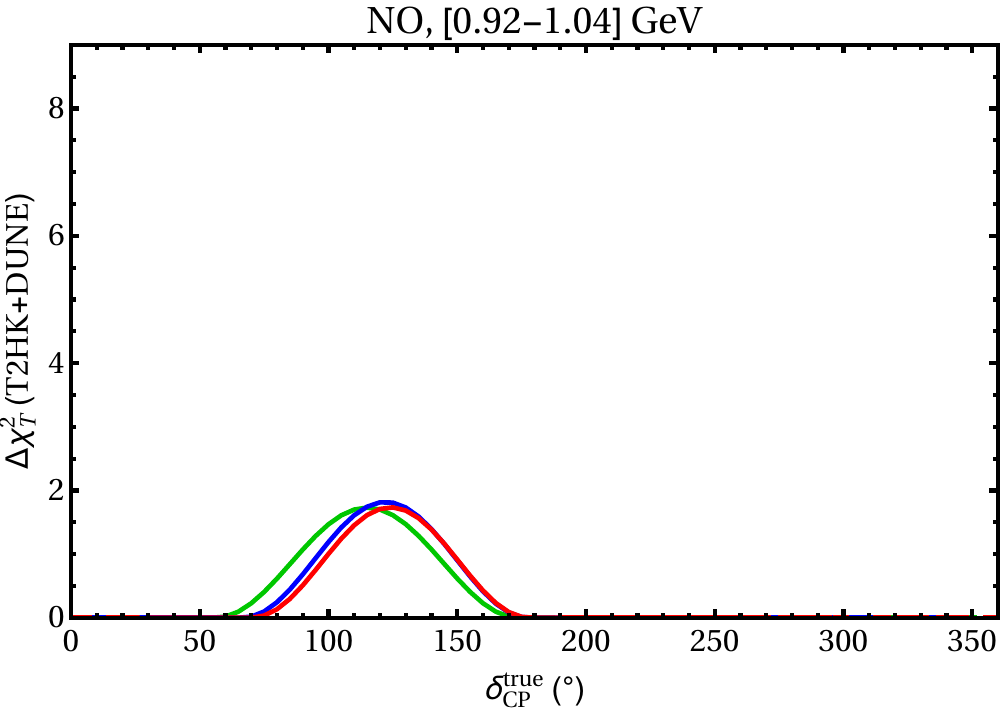}
    \includegraphics[width=0.5\linewidth]{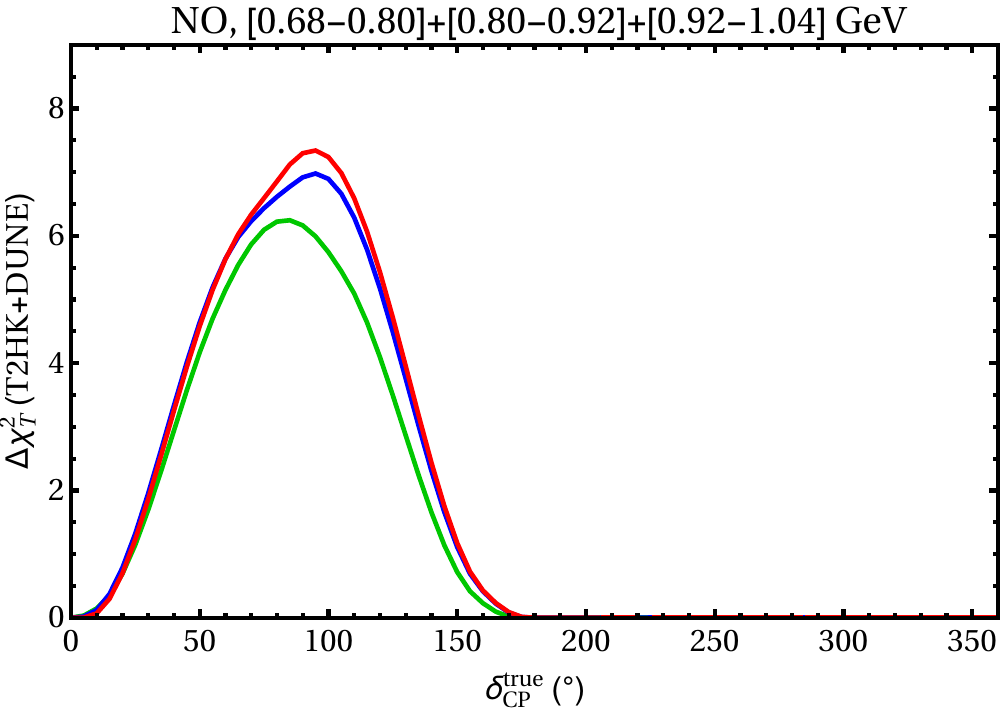}
    \mycaption{T2HK + DUNE sensitivity to T violation as a function of true CP phase $\delta_{\rm CP}^{\rm true}$. The top-right panel corresponds to the predicted sensitive energy window [0.80-0.92]~GeV, \cref{eq:bin}, top-left and bottom-left show its lower and upper neighboring bins; in the bottom-right panel we show the combination of the three bins. Green curves correspond to the default energy resolution according to \cref{eq:resolution}; for blue and red curves we re-scale the width of the resolutions globally by a factor 0.5 and 0.2, respectively. 
    We assume exposures of 608 (840)~kt~MW~yr for T2HK (DUNE) in the neutrino mode.}
    \label{fig:sens-deltaCP}
\end{figure}

We now present the results of our numerical sensitivity calculations. \Cref{fig:sens-deltaCP} shows the value of the $\chi^2$ statistic \cref{eq:chi2T} as a function of the assumed true value of the 3-flavour CP phase for our default T2HK and DUNE configurations. 
The final sensitivity is displayed in the bottom-right panel, which includes the full energy range relevant for the test. We observe that there is only sensitivity in the range $0< \delta_{\rm CP} < 180^\circ$. The reason is because for this figure we assume only the neutrino beam mode. For $180^\circ < \delta_{\rm CP} < 360^\circ$, antineutrinos offer sensitivity in principle \cite{Schwetz:2021cuj}, which however, turns out to be very poor for realistic experimental configurations, as we will discuss below in \cref{sec:anti}. Therefore, we focus on neutrino running, in which case we obtain sensitivities slightly below $3\sigma$ for $\delta_{\rm CP} \simeq 90^\circ$. 

The top-right panel of \cref{fig:sens-deltaCP} shows $\Delta\chi^2_T$ using only the energy bin from \cref{eq:bin}, where we expected sensitivity for normal ordering and neutrino mode, whereas top-left and bottom-left panels correspond to the neighbouring bins below and above. According to our model-independent approach, we independently minimize with respect to the $c_i$ coefficients in each bin. Clearly, the bulk of the sensitivity is provided by the predicted energy bin, confirming our analytical estimates. The neighbouring bins do provide minor sensitivity, mostly because of smearing effects due to finite energy resolution, which we are going to discuss in more detail now. 

We consider a Gaussian detector resolution with $\sigma= \alpha E_\nu + \beta \sqrt{E_\nu} + \gamma$, where $E_\nu$ is the neutrino energy in GeV. We adopt the following default configuration (units are GeV) 
\begin{equation} \label{eq:resolution}   
 (\alpha,\beta,\gamma)=
 \left\{
 \begin{array}{l@{\qquad}l}
 (0.12,0.07,0.0) & \text{T2HK neutrinno} \\ 
 (0.12,0.0,0.09) & \text{T2HK antineutrino} \\
 (0.045,0.001,0.048) & \text{DUNE neutrino} \\ 
 (0.026,0.001,0.085) & \text{DUNE antineutrino}
 \end{array}
 \right.
\end{equation}
For T2HK these numbers have been chosen in order to match the results provided  in the design report~\cite{Hyper-Kamiokande:2018ofw}; 
for DUNE we adopt an improved energy resolution based on \cite{Friedland:2018vry,Chatterjee:2021wac}. For $E_\nu= 0.86$~GeV, these assumptions imply a neutrino energy resolution of about 19\% for T2HK and 10\% for DUNE. \Cref{fig:sens-deltaCP} shows the sensitivity for this default assumption as green curves as well as the impact of improved energy resolutions, multiplying the numbers from \cref{eq:resolution} by a factor 0.5 (blue curves) or 0.2 (red curves).  We see that the main impact of improving the energy resolution is to shift sensitivity from the lower energy bin (sensitivity decreases with improved resolution) towards the central bin (sensitivity improves with resolution). This behaviour again supports our analytical arguments, that the sensitivity is dominated by true neutrino energies corresponding to the central bin from 0.8 to 0.92~GeV.  

\begin{figure}[t]
    \centering
    \includegraphics[width=0.7\textwidth]{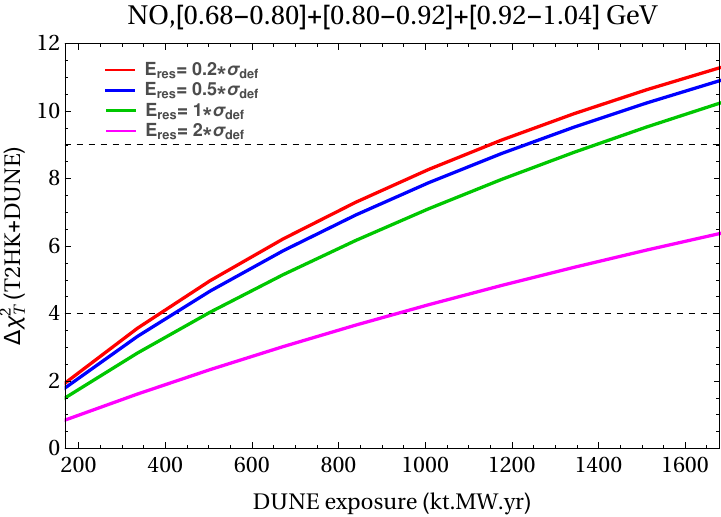}
    \mycaption{$\Delta\chi^2_T$ as a function of the DUNE neutrino exposure for true $\delta_{\rm CP} = 90^\circ$ summing the three relevant energy bins. The T2HK exposure is kept fixed at 608~kt~MW~yr. Different curves correspond to different assumptions on the neutrino energy resolutions. The green curve represents our default resolution according to \cref{eq:resolution}, for the red (blue) curve the resolutions for both, T2HK and DUNE have been re-scaled by a factor of 0.2 (0.5), while for the magenta curve only the DUNE resolution has been re-scaled by a factor 2.}
    \label{fig:exposure}
\end{figure}

Let us now study the interplay of energy resolution and exposure. The estimates in \cref{sec:sens-estimate} suggest that the statistical uncertainty is dominated by DUNE under our default assumptions of 608 (840)~kt~MW~yr for T2HK (DUNE), see \cref{eq:stat-error}. \Cref{fig:exposure} shows the T violation sensitivity for $\delta_{\rm CP}^{\rm tr} = 90^\circ$ as a function of the DUNE exposure in neutrino mode for different assumptions on the energy resolutions in both experiments. We see that an improved energy reconstruction can somewhat reduce the required exposure to reach a certain sensitivity. Note that our default assumption for DUNE according to \cref{eq:resolution} corresponds already to an improved resolution according to \cite{Friedland:2018vry}. The magenta curve in \cref{fig:exposure} shows the sensitivity for a DUNE resolution reduced by a factor two compared to \cref{eq:resolution}, which corresponds to good accuracy to the value from the DUNE TDR \cite{DUNE:2020jqi,DUNE:2021cuw}. We see that the improved reconstruction is essential to obtain good sensitivity with reasonable exposures.

\subsection{Zero-distance effect and prior on oscillation frequencies} \label{sec:ND}

\begin{figure}[t]
    \centering
    \includegraphics[width=0.7\linewidth]{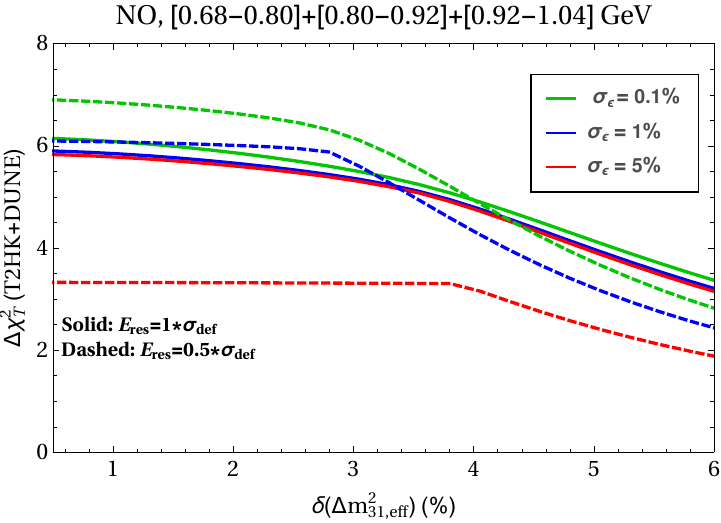}
    \mycaption{Sensitivity to T violation for $\delta_{\rm CP}^{\rm true} = 90^\circ$ as a function of the prior on the effective mass-squared difference in matter, $\Delta m^2_{31,\rm eff}$, for different assumptions on the near detector constraint on the zero-distance effect, $\sigma_\epsilon = 0.1\%, 1\%, 5\%$ for the green, blue, red curves, respectively. For the solid curves we assume the default energy resolution from \cref{eq:resolution}, for the dashed curves the energy resolution is re-scaled by a factor 0.5 for both experiments. Exposures have been set to our default assumptions.}
    \label{fig:prior-ND}
\end{figure}

According to assumption ($iv$) stated in \cref{sec:framework}, the oscillation frequencies $\lambda_i-\lambda_j$ differ only slightly from the corresponding standard $3\nu$ case. In our default analysis we have fixed the oscillation phases to the value given in \cref{eq:phi}, where $\Delta m^2_{ij,\rm eff}$ is the standard effective mass-squared difference in matter for our assumed constant matter density and the central neutrino energy of the relevant energy bin. In order to quantify the accuracy with which $\Delta m^2_{ij,\rm eff}$ has to be known we treat $\Delta m^2_{31,\rm eff}$ as an additional free parameter in the fit constrained by a Gaussian prior whose uncertainty is shown on the horizontal axis in \cref{fig:prior-ND} centered around the standard model value. The interpretation of this prior width is two-fold: first, it quantifies the assumption that new-physics contributions to the oscillation frequencies have to be ``small'' and second, it provides a measure of the precision needed on the standard mass-squared differences from additional data, e.g., from the disappearance data of the same experiments (see \cite{Schwetz:2021cuj}) or, under more model-dependence, also from external data on the oscillation frequencies. We have checked that our results are completely insensitive to uncertainties on $\Delta m^2_{21,\rm eff}$ up to $6\%$. Therefore, fixing $\Delta m^2_{21,\rm eff}$ to its SM value is a very good assumption and we focus our discussion below on the uncertainty of $\Delta m^2_{31,\rm eff}$.

Second, we want to study the impact of the near detector constraint $\sigma_\epsilon$.
The ``near detector'' prior constrains the deviation from unitarity parametrized by the parameter $\epsilon = \sum_i c_i$, see \cref{eq:epsilon,eq:chi2k}. As discussed in \cref{sec:analysis} and \cref{app:zero-distance} this constraint corresponds to an effective constraint which emerges from a combination of genuine new-physics non-unitarity as well as the actual near-detector measurements of the considered experiments. We treat $\sigma_\epsilon$ as an effective parameter which we set to 0.1\% in our default analysis, whereas in \cref{fig:prior-ND} we study the dependence of the sensitivity on this assumption. 

\Cref{fig:prior-ND} shows a non-trivial interplay of the near-detector constrains (curves with different colors), the energy resolution (solid curves: default assumption, dashed curves: improved by a factor 0.5), and the oscillation frequency prior. A crucial role is played by the quantity $\delta_0$ defined in \cref{eq:delta0}, which multiplies the non-unitarity factor $\epsilon^2$ in the observable $X_T$, see \cref{eq:XT}. The value of $\delta_0$ is determined by the oscillation frequencies and therefore depends on energy. Hence, effectively we have to average $\delta_0$ over the width of the considered energy bin and fold it with the resolution function. It turns out that the energy-averaged value of $\delta_0$ monotonically decreases from about $-0.09$ for our default energy resolution to $-0.21$ for a resolution improved by a factor 0.2. Hence, we expect that the impact of the zero-distance effect becomes more important for better energy resolution. This is the behaviour visible in \cref{fig:prior-ND}: while the solid curves are rather insensitive to the value of the near-detector constraint even up to $\sigma_\epsilon = 5\%$, we see a rather strong dependence on $\sigma_\epsilon$ for the improved energy resolution (dashed curves) and the sensitivity degrades significantly for uncertainties $\sigma_\epsilon \gtrsim 1\%$. This behaviour is consistent with the estimate in \cref{eq:stat-error}. Note that for $\sigma_\epsilon = 5\%$ the better resolution leads to a worse sensitivity, which is a manifestation of the larger value of $|\delta_0|$ due to the non-trivial energy averaging, see also \cref{eq:stat-error1}. 

This effect is further entangled with the uncertainty on $\Delta m^2_{31,\rm eff}$. A large uncertainty on $\Delta m^2_{31,\rm eff}$ has a similar effect as the energy smearing. The conditions \cref{eq:cond1,eq:cond2} which have to be fulfilled in order for the observable $X_T$ being sensitive to T violation depend on the oscillation frequency. Sizeable uncertainties in the frequencies allow to change the values such that the conditions 
\cref{eq:cond1,eq:cond2} no longer hold in certain regions in the relevant energy range, which degrades the over-all sensitivity. We conclude from \cref{fig:prior-ND} that the precision on the oscillation frequency should be better than about 2\% before the sensitivity degrades significantly.

For the improved energy resolution we notice also a flat plateau when increasing the frequency prior for $\sigma_\epsilon \gtrsim 1\%$ (blue and red dashed curves). The origin of this behaviour is related to multiple minima in the $\chi^2$ in the $c_i$ space which appear when allowing for non-unitarity. For large enough $\sigma_\epsilon$ a minimum appears which has a best fit point of $\Delta m^2_{31,\rm eff}$ very close to its central value, and is therefore independent of the assumed prior width. The curve starts to deviate from the plateau when another minimum becomes the global one, which then is affected by the $\Delta m^2_{31,\rm eff}$ prior. 

To summarize, for uncertainties on the oscillation frequency and/or the zero-distance effect $\gtrsim 1\%$ a complicated interplay appears, leading to counter-intuitive behaviour with respect to energy resolution because of  non-trivial energy-averaging effects. For robust results of the proposed T violation test, constraints on the zero-distance effect and on the oscillation frequency better than $\simeq 1\%$ are desirable.

\subsection{Antineutrino beam mode} \label{sec:anti}

Before concluding let us briefly comment on the sensitivity of the antineutrino beam mode. Antineutrinos can in principle provide sensitivity for $180^\circ < \delta_{\rm CP} < 360^\circ$~\cite{Schwetz:2021cuj} as for these values of $\delta_{\rm CP}$, $X_T$ is negative for standard oscillations. However, in the experimental setup considered here, the sensitivity is only very poor for two main reasons. First, event numbers for DUNE in the relevant energy bin are very small, and second, there is a large neutrino component in the antineutrino beam, which leads to a dilution of the T violation effect.

\begin{figure}[t]
    \centering
    \includegraphics[width=0.7\linewidth]{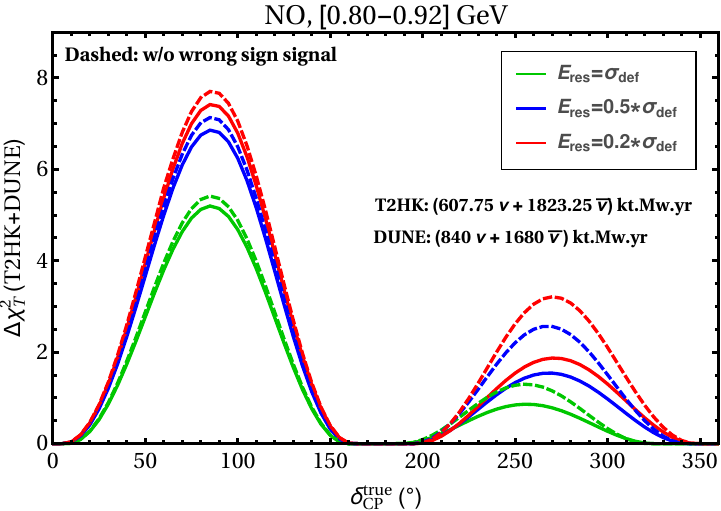}
    \mycaption{Sensitivity to T violation as a function of true CP phase $\delta_{\rm CP}^{\rm true}$ combining 608~kt~MW~yr (T2HK) and 840~kt~MW~yr (DUNE) exposure in the neutrino beam mode with 1823~kt~MW~yr (T2HK) and 1680~kt~MW~yr (DUNE) in the antineutrino beam mode. We show results for the energy bin [0.80-0.92]~GeV. Green curves correspond to the default energy resolution according to \cref{eq:resolution}; for blue and red curves we re-scale the width of the resolutions globally by a factor 0.5 and 0.2, respectively. For the dashed curves we remove the ``wrong sign signal'' due to the $\bar\nu_\mu$ ($\nu_\mu$) beam component in the neutrino (antineutrino) beam mode.}
    \label{fig:anti}
\end{figure}

In \cref{fig:anti} we show the sensitivity by analysing neutrino and antineutrino exposure simultaneously, with independent $c_i$ coefficients. In order to obtain a meaning full number of events in DUNE we assume the very large exposure of 1680~kt~MW~yr for antineutrinos, which gives only 6 signal events for $\delta_{\rm CP} = 270^\circ$, c.f.\ \cref{fig:event-spectra}. It is apparent from \cref{fig:anti} that in the region $180^\circ < \delta_{\rm CP} < 360^\circ$ only limited sensitivity can be achieved, despite the (unrealistically) large DUNE exposure. 

Another reason is the relatively large neutrino component in the ``antineutrino'' flux mode. Indeed, in the relevant energy bin for DUNE we expect even more ``wrong-sign'' neutrino events than antineutrino events: from \cref{fig:event-spectra} we find 12 neutrino versus 6 antineutrino events. Also for T2HK the wrong-sign component is sizable with 26 neutrino versus 140 antineutrino events. Note that in our model-independent approach, neutrino and antineutrino transition probabilities are governed by a different set of $c_i$ coefficients. For this reason, a meaning-full analysis of antineutrino beam running is only possible together with data from the neutrino mode, in order to constrain both set of $c_i$'s.\footnote{Note that in the neutrino beam mode, the wrong-sign component is very small: 3 versus 180 signal events for T2HK and 1 versus 18 for DUNE for our default exposures. Therefore, it is sensible to consider neutrino data independently. For any reasonable values of $c_i^{\bar\nu}$ the contribution of the antineutrinos will be negligible.} This is further illustrated by the dashed curves in \cref{fig:anti}, for which we switch off the ``wrong-sign'' beam components for illustration purposes. However, even in this hypothetical case we can achieve at best sensitivities up to $2\sigma$ due to the small event numbers in DUNE.

\section{Summary and discussion}\label{sec:conclusion}

In this work we have pointed out that by combining measurements of the $\nu_\mu\to\nu_e$ transition probabilities at T2HK and DUNE the fundamental time reversal symmetry can be tested model-independently. 
We have proposed a simple test, based on the observable $X_T$ introduced in \cref{eq:XTobs}, which is the difference of the transition probabilities at DUNE, at T2HK and at zero-distance. An observation of $X_T < 0$ around a neutrino energy of $E_\nu \simeq 0.86$~GeV 
implies violation of the T symmetry. If a sufficiently strong constraint on the zero-distance effect is available, $X_T < 0$ just implies a probability at the DUNE baseline smaller than at the T2HK baseline. 

Under the assumption of oscillation frequencies approximately as in the three-flavour standard neutrino case, we have searched for possible two-baseline and neutrino energy combinations and identified the T2HK/DUNE combination at $E_\nu \simeq 0.86$~GeV as a rather unique spot where the test can be applied. Another potentially interesting region has been found for baselines around 2000 to 3000~km combined with DUNE at neutrino energies around 1.4~GeV; a more detailed investigation of which is left for future work. 

We have performed numerical studies based on GLoBES to investigate the experimental requirements for the DUNE/T2HK test, using T violation due to the standard three-flavour Dirac CP phase as example. The most important conclusions are the following:
\begin{itemize}
\item The sensitive energy interval is located close to the maximum of the appearance event spectrum for T2HK, but for DUNE it appears in the low energy tail of the event spectrum and suffers from limited statistics. Therefore, possibilities to increase event numbers in the sub-GeV region for DUNE are required to obtain sufficient statistical precision for the test. With current DUNE beam and detector configuration, run times of order 10 years in neutrino mode are required.
\item A somewhat improved energy resolution is important to reach good sensitivities to T violation. For DUNE, at least a resolution of around 10\% at $E_\nu = 1$~GeV (as suggested e.g., in \cite{Friedland:2018vry}) is required, whereas any further improvement in energy reconstruction for both, T2HK and DUNE would increase the T violation sensitivity.
\item A sufficiently strong constraint on $\nu_\mu\to\nu_e$ transitions at zero distance is required, at least at the $\lesssim 1\%$ level. As this requirement is typically looser than the allowed size of non-unitarity effects from generic new physics \cite{Ellis:2020hus,Hu:2020oba,Forero:2021azc,Blennow:2023mqx}, this can be interpreted as the required near-detector measurement precision to constrain the impact of flux uncertainties.
\item In the antineutrino beam mode we do not reach relevant sensitivity to T violation. The two main reasons are the limited statistics in the sub-GeV region in DUNE as well as the large neutrino component in the antineutrino flux mode, which leads to a dilution of the T violation effect. To explore our proposed test for antineutrinos substantial improvements with respect to the current experimental configuration for DUNE are necessary.
\end{itemize}

Let us summarize in which sense our approach is model-independent: our analysis is very general in terms of the effective mixing parameters relating the flavour states and the eigenstates of the propagation Hamiltonian. We treat the $c_i$ coefficients (see \cref{eq:P,eq:mixing}) completely unconstrained in each energy bin and separate for neutrinos and antineutrinos. The strongest assumption is that only two oscillation frequencies are relevant, and that they are numerically close (within $\lesssim 2\%)$ to the standard oscillation frequencies in matter assuming the Standard Model matter effect. Note that this assumptions does not only involve the (rather precisely measured) vacuum mass-squared differences but also the impact of non-standard interactions potentially modifying the matter effect. The oscillation frequencies determine the conditions \cref{eq:cond1,eq:cond2} on the neutrino energy and the two baselines which have to be fulfilled in order for the observable $X_T$ being sensitive to T violation. Sizeable uncertainties in the frequencies lead to a complicated interplay of energy resolution and zero-distance effects, affecting the sensitivity to T violation in a non-trivial way, see discussion in \cref{sec:ND}.

To conclude, we encourage the neutrino oscillation community to take our proposal into consideration, as it offers a unique possibility to search for fundamental T violation in neutrino oscillations in a rather direct and model-independent way. Under the well founded assumption of CPT conservation, this would allow for an independent test of the CP symmetry and offer complementary information on the symmetries of the fundamental theory of leptons.

\subsection*{Acknowledgements}

We thank Evgeny Akhmedov for discussions about the interpretation of the time reversal transformation in the context of neutrino oscillations. The work of S.S.C.\ is funded by the Deutsche Forschungsgemeinschaft (DFG, German Research Foundation) –- project number 510963981. 
K.S.\ acknowledges the financial support received from the Bi-nationally Supervised Doctoral Degrees program of the German Academic Exchange Service (DAAD). K.S.\ is also very much thankful to Thomas Schwetz-Mangold for the warm hospitality and research facilities provided at the Institute for Astroparticle Physics, KIT, Germany. 
This work has been supported by the European Union’s Framework Programme for Research and Innovation Horizon 2020 under grant H2020-MSCA-ITN-2019/860881-HIDDeN.

\begin{appendix}

\section{Time reversal in the QFT formalism for neutrino oscillations}
\label{sec:QFT}

The test for T violation considered in this work is based on the property described in \cref{eq:TL}, which states that time reversal is equivalent to the operation $L\to -L$, which actually corresponds to inversion of space. While the equivalence of these transformations is explicit in the standard formula for oscillation probabilities, it remains unclear on a more fundamental level in what sense the transformation $L\to-L$ can be considered as ``time reversal''. The origin of this association emerges as follows. Fundamentally the evolution equation in quantum mechanics, i.e., the Schr\"odinger equation is expressed as an equation in time. However, the neutrino flavour system is special in the sense, that source and detector are macroscopically separated. Taking into account that source an detector particles necessarily need to be localised and described by wave packets, neutrino propagation can be effectively described by the approximation $x\approx vt \approx t$~\cite{Akhmedov:2010ms}. This {\it a posteriori} result can then be used to re-write the Schr\"odinger equation in terms of space, leading to the usuall oscillation probability in terms of distance.

In this appendix we provide a short derivation of the equivalence of \cref{eq:T,eq:TL} based on the quantum-field theoretical (QFT) approach to neutrino oscillations~\cite{Rich:1993wu,Giunti:1993se,Grimus:1996av,Kiers:1997pe,Beuthe:2001rc}. We consider the time reversal transformation $\mathcal{T}$ of the neutrino oscillation amplitude in the QFT formalism and show that the well-known relation
\begin{equation}\label{eq:Tappendix}
  \mathcal{T}[P_{\alpha\to\beta}(L)] = P_{\beta\to\alpha}(L) = P_{\alpha\to\beta}(-L) \,. 
\end{equation}
can be derived from the time reversed transition amplitude in QFT.

\bigskip
\textbf{Neutrino oscillations in the QFT formalism.}
Let us consider the neutrino oscillation process
where a neutrino of flavour $\alpha$ is produced at time $t_P$ at the location $\mathbf{x}_A$ and
detected as flavour $\beta$ at time $t_D$ at the location $\mathbf{x}_B$:
\begin{equation}
  (\nu_\alpha,t_P,\mathbf{x}_A) \quad\to\quad
  (\nu_\beta,t_D,\mathbf{x}_B) \,, \label{eq:process}
\end{equation}
We describe this process using the QFT formalism, following the notation of \cite{Krueger:2023skk}.
We depart from Eq.~(3.4) of \cite{Krueger:2023skk} for the transition amplitude of the process \cref{eq:process}, which we rewrite in the following way:
\begin{align}
  i\mathcal{A}_{\alpha\to\beta}(T,L) =& 
  \frac{1}{4\pi L}
  \left(\prod_{i=A,B,f} \mathcal{N}_i\right)
  \frac{\pi^4}{\sigma_{pP}^3\sigma_{EP}\sigma_{pD}^3\sigma_{ED}}
  i\mathcal{M}_\alpha^P  i\mathcal{M}_\beta^D \,
  i\mathcal{A}_{\alpha\to\beta}^{\rm red}(T,L) \,, \label{eq:A}\\
  i\mathcal{A}_{\alpha\to\beta}^{\rm red}(T,L) =& 
  \sum_j U_{\alpha j}^*U_{\beta j}
  \int \frac{dp^0}{2\pi}
  \exp\left[ -ip^0T + ip_jL -f_j(p^0) \right] \,. \label{eq:Ared}
\end{align}
where , 
\begin{align} \label{eq:defs}
  T = t_D - t_P\,,\quad
  L = |\mathbf{x}_B - \mathbf{x}_A| \,,\quad
  p_j = \sqrt{(p^0)^2 - m_j^2} \,.
\end{align}
The real function $f_j(p^0)$ takes into account the approximate energy conservation according to the wave packet spread of external particles. 
Here we have reduced the $d^4p$ integral over the neutrino 4-momentum
by performing the $d^3p$ integral 
using the Grimus-Stockinger theorem \cite{Grimus:1996av}. This allows
us to take into account the macroscopic separation of source and
detector such that neutrinos go on-shell.

In \cref{eq:A} we have factorized the total amplitude into the amplitudes
  $\mathcal{M}_\alpha^P$ and  $\mathcal{M}_\beta^D$ describing the production and detection processes, respectively, and the amplitude $\mathcal{A}_{\alpha\to\beta}^{\rm red}(T,L)$ responsible for the flavour oscillation process. The assumptions are that 
$\mathcal{M}_\alpha^P$ and  $\mathcal{M}_\beta^D$ are independent of the neutrino mass $j$ and are sufficiently slow functions of the neutrino energy, such that they can be pulled out of the $dp^0$ integral. In \cref{eq:Ared} we have isolated the reduced amplitude, which describes the flavour transition and which we will use below to study the time reversal operation. 

In the above approximation, one can identify an oscillation probability, independent of the production and detection process, which will be proportional to $|\mathcal{A}_{\alpha\to\beta}^{\rm red}(T,L)|^2$. However, in general the time of neutrino production is not observed (or not known with sufficient precision), such that an average over the production time $t_P$ has to be performed, or equivalently an integral over $T$:
\begin{align}
  P_{\alpha\to\beta}(L) &\propto \overline{|\mathcal{A}_{\alpha\to\beta}^{\rm red}(T,L)|^2}
  \propto \int dT |\mathcal{A}_{\alpha\to\beta}^{\rm red}(T,L)|^2 \label{eq:P1}\\
  &\propto
  \sum_{jk} U_{\alpha j}^*U_{\beta j}U_{\alpha k}U_{\beta k}^*
  \int dp^0 \exp\left[ i\frac{\Delta m^2_{kj}L}{2p^0} - f_j(p^0)-f_k(p^0) \right] \label{eq:P2}\\
  &\approx
  \sum_{jk} U_{\alpha j}^*U_{\beta j}U_{\alpha k}U_{\beta k}^*
  \exp\left[ i\frac{\Delta m^2_{kj}L}{2 E_\nu} \right] \mathcal{D}_{jk} \,. \label{eq:P3}
 \end{align}
In the step from the first to the second line we have used that the $T$-integral gives a $\delta$-function in $p^0$ and we have taken into account that neutrino masses are small, $m_j \ll p^0$, and
expand the square root in the neutrino momenta as $p_j \approx p^0 - m_j^2/(2p^0)$.
In the last line, $E_\nu$ is an effective neutrino energy and the coefficient $\mathcal{D}_{jk}$ takes into account possible decoherence effects emerging from the $dp^0$ integral and the function $f_j(p^0)$. For all cases of interest we have $\mathcal{D}_{jk} \approx 1$ to very good approximation and 
\cref{eq:P3} corresponds to the ``standard'' oscillation formula.

\textbf{The time reversal transformation.}
Let us now consider the time reversal transformation $\mathcal{T}$ and study how it affects the final oscillation probability by departing from the QFT amplitude \cref{eq:Ared}. The time reversed process to the one considered in \cref{eq:process} is the following:
\begin{equation}
  \mathcal{T}[(\nu_\alpha,t_P,\mathbf{x}_A) \to
    (\nu_\beta,t_D,\mathbf{x}_B)]
   = [(\nu_\beta,t_P,\mathbf{x}_B) \to
    (\nu_\alpha,t_D,\mathbf{x}_A)]
  \,, \label{eq:Tprocess}
\end{equation}
i.e., we consider a neutrino of flavour $\beta$ produced at a time $t_P$ at the position $\mathbf{x}_B$ and a neutrino of flavour $\alpha$ detected at time $t_D$ at the position $\mathbf{x}_A$. Hence, if T is conserved, this would be the process seen if a hypothetical video recording of the original process \cref{eq:process} would be run backwards.\footnote{Note that this differs from the standard time reversal operation in quantum physics, which would reverse also the helicity of the neutrino states, whereas here we want to maintain neutrino helicity.} The amplitude for the time reversed process therefore is 
\begin{align}
  \mathcal{T}[i\mathcal{A}_{\alpha\to\beta}^{\rm red}(T,L)]  & =
  i\mathcal{A}_{\beta\to\alpha}^{\rm red}(T,L) =
  \sum_j U_{\beta j}^*U_{\alpha j}
  \int \frac{dp^0}{2\pi}
  \exp\left[ -ip^0T + ip_jL -f_j(p^0) \right]
    \label{eq:AredT1}\\
  & = [i\mathcal{A}_{\alpha\to\beta}^{\rm red}(-T,-L)]^* \,.
  \label{eq:AredT2}
\end{align}
We note that $L$ is defined as modulus in \cref{eq:defs} and remains positive although we exchange $\mathbf{x}_A$ and $\mathbf{x}_B$.\footnote{The term $p_jL$ originates from a term
$\mathbf{p} \cdot (\mathbf{x}_A - \mathbf{x}_B)$, and the Grimus-Stockinger theorem makes sure that the $d^3p$ integral picks momenta aligned with the vector pointing from the production to the detection point, such that $\mathbf{p} \cdot (\mathbf{x}_A - \mathbf{x}_B) \to p_j L > 0$.}

Using the relation between the amplitude and the probability in \cref{eq:P1,eq:P2,eq:P3} the relation \cref{eq:Tappendix} follows immediately from \cref{eq:AredT1,eq:AredT2}. Hence, we confirm that also in the QFT formalism the time reversal transformation corresponds to
\begin{enumerate}
\item swaping initial and final flavour of the probability, or equivalently
\item applying the transformation $L\to -L$ in the probability.  
\end{enumerate}
%
The above argument applies in a straight forward way to oscillations in matter
in the limit of approximately constant matter potential \cite{Akhmedov:2012mk}
as well as to the non-standard scenarios considered in \cite{Schwetz:2021cuj,Schwetz:2021thj}.

\section{Constraints on the zero-distance effect}
\label{app:zero-distance}

Let us first assume that the zero-distance effect $\epsilon$, \cref{eq:ND}, is constrained only by the near detectors of the two experiments used to construct the observable $X_T$. The covariance matrix for two measurements for the transition probabilities at the far and near detectors of one experiment is then given by
\begin{equation}
    S_i = \left(\begin{array}{cc}
       \sigma_{i,\rm f}^2 & 0\\
       0 & \sigma_{i,\rm n}^2 
    \end{array}\right)
    + \sigma_{i,\rm c}^2 \left(\begin{array}{cc}
    1 & 1 \\ 1 & 1
    \end{array}\right) \,,
\end{equation}
where $\sigma_{i,\rm f}$ ($\sigma_{i,\rm n}$) are the statistical and uncorrelated systematic errors of the far (near) detector measurements and $\sigma_{i,\rm c}$ is a fully correlated error including for instance flux and cross section uncertainties as well as common systematics. The index $i=1,2$ labels the two experiments. Note that $\sigma_{i,\rm n}$ is the uncertainty on the $P_{\nu_\mu\to\nu_e}$ transition probability at the near detector. Hence, it will be dominated by the statistical and systematic errors of the $\nu_e$ beam background. 

We can calculate the combined covariance matrix of the three quantities $P(L_1),P(L_2), \epsilon^2 \equiv P(L=0)$ by summing the inverse covariance matrices of the two experiments.
Straight-forward application of error propagation allows then to calculate the total uncertainty on $X_T$. We write
\begin{align}
    \sigma_{X_T}^2 = \sigma_{1,\rm f}^2 + \sigma_{2,\rm f}^2 + \delta_0^2 \sigma_{\epsilon,\rm eff}^2 \,, \label{eq:sigmaX}
\end{align}
where $\sigma_{\epsilon,\rm eff}$ is the effective ``near detector'' constraint on $\epsilon^2$ used in our numerical simulations. In terms of the uncertainties introduced above it is given by 
\begin{align}
    \delta_0^2 \sigma_{\epsilon,\rm eff}^2 = 
    \frac{\delta_0^2 (\sigma_{1,\rm n}^2 + \sigma_{1,\rm c}^2)(\sigma_{2,\rm n}^2 + \sigma_{2,\rm c}^2) +
    (\sigma_{1,\rm n}^2 + \sigma_{2,\rm n}^2)(\sigma_{1,\rm c}^2 + \sigma_{2,\rm c}^2) + 2\delta_0(\sigma_{1,\rm n}^2  \sigma_{2,\rm c}^2 - \sigma_{2,\rm n}^2\sigma_{1,\rm c}^2)   
    }{\sigma_{1,\rm n}^2 + \sigma_{2,\rm n}^2 + \sigma_{1,\rm c}^2 + \sigma_{2,\rm c}^2} \,.
    \label{eq:sig_eff1}
\end{align}
Assuming that correlated errors (such as flux uncertainties) are much larger than statistical and uncorrelated errors at the near detectors, the effective uncertainty will be dominated by the correlated errors:
\begin{align}
    \sigma_{\epsilon,\rm eff}^2 \to \frac{\sigma_{1,\rm c}^2\sigma_{2,\rm c}^2}{\sigma_{1,\rm c}^2+\sigma_{2,\rm c}^2} 
    \quad\text{for}\quad \sigma_{i,\rm n} \ll \sigma_{i,\rm c} \,.
\end{align}
Hence, we see that in this most general framework, correlated uncertainties will not cancel by the near/far combination and will actually dominate the effective uncertainty on $\epsilon$. 

This is the most conservative case, when only information on the $\nu_\mu\to\nu_e$ transitions is used. However, typically we can assume that additional external constraints are available:
\begin{enumerate}
    \item Fully correlated uncertainties in an experiment are usually constrained by additional near detector measurements, in particular the $\nu_\mu$ flux measurement. Under modest model-dependence, we can therefore assume that $\sigma_{i,\rm c}$ is of the order of the near detector measurement uncertainty in the $\nu_\mu$ channel, which typically is much smaller than the one in the $\nu_e$ channel. In this limit we will have
 \begin{align}
    \sigma_{\epsilon,\rm eff}^2 \to \frac{\sigma_{1,\rm n}^2\sigma_{2,\rm n}^2}{\sigma_{1,\rm n}^2+\sigma_{2,\rm n}^2} 
    \quad\text{for}\quad \sigma_{i,\rm c} \ll \sigma_{i,\rm n} \,,
\end{align}
i.e., the zero-distance transition probability 
$\epsilon^2$ is indeed constrained by the 
$\nu_\mu\to\nu_e$ measurement at the near detector (which still would be dominated by the uncertainty on the intrinsic $\nu_e$ beam background).
\item
Under modest model-dependent assumptions we can apply external constraints on 
the zero-distance effect, which set tight limits on the ND probability $\epsilon^2$, see the discussion in \cref{sec:analysis}. Let us denote the uncertainty of such external constraints by $\sigma_{\epsilon,\rm ext}$. In this case we obtain for the effective uncertainty
\begin{align}
    \delta_0^2 \sigma_{\epsilon,\rm eff}^2 = 
    \frac{\sigma_{\epsilon,\rm ext}^2 N + (\sigma_{1,\rm n}^2 + \sigma_{2,\rm n}^2)  \sigma_{1,\rm c}^2  \sigma_{2,\rm c}^2 + \sigma_{1,\rm n}^2  \sigma_{2,\rm n}^2  (\sigma_{1,\rm c}^2 + \sigma_{2,\rm c}^2)}
    {\sigma_{\epsilon,\rm ext}^2 D + \sigma_{1,\rm n}^2  \sigma_{2,\rm n}^2 +  \sigma_{1,\rm c}^2  \sigma_{2,\rm c}^2 + \sigma_{1,\rm n}^2  \sigma_{2,\rm c}^2  + \sigma_{1,\rm c}^2  \sigma_{2,\rm n}^2} \,,
    \label{eq:sig_eff2}
\end{align}
where $N$ ($D$) is the numerator (denominator) of \cref{eq:sig_eff1}. In the limit of small $\sigma_{\epsilon,\rm ext}$ we obtain
\begin{align}
\delta_0^2 \sigma_{\epsilon,\rm eff}^2 \to 
\left\{\begin{array}{l}
    \delta_0^2 \sigma_{\epsilon,\rm ext}^2
    + \sigma_{1,\rm n}^2 + \sigma_{2,\rm n}^2 
    \quad\text{for}\quad \sigma_{\epsilon,\rm ext},\, \sigma_{i,\rm n} \ll \sigma_{i,\rm c} \,, \\
    \delta_0^2 \sigma_{\epsilon,\rm ext}^2
    + \sigma_{1,\rm c}^2 + \sigma_{2,\rm c}^2 
    \,\quad\text{for}\quad \sigma_{\epsilon,\rm ext},\, \sigma_{i,\rm c} \ll \sigma_{i,\rm n} \,.
    \end{array}
\right.
\end{align}
Hence, the final uncertainty is set by the {\it smaller} of $\sigma_{i,\rm c}$ and $\sigma_{i,\rm n}$. 
The total uncertainty on the observable $X_T$ is then typically dominated by the far-detector uncertainties, see \cref{eq:sigmaX}.
\end{enumerate}

In summary, the zero-distance constraint $\sigma_\epsilon$ introduced in \cref{eq:chi2k} in \cref{sec:simulation} should be understood as $\sigma_{\epsilon,\rm eff}$ as given in \cref{eq:sig_eff2}. It emerges as a combination of uncorrelated near detector uncertainties, near-far correlated errors, and external constraints on the zero-distance effect.

\end{appendix}

\bibliographystyle{JHEP_improved}
\bibliography{./refs}

\end{document}